\documentclass[journal]{IEEEtran}
\usepackage{graphicx}
\usepackage{amstext}
\usepackage{algorithm}
\usepackage{algorithmic}
\usepackage{mathrsfs}
\usepackage{amssymb}
\usepackage{amsmath}
\usepackage{bm}
\allowdisplaybreaks[4]
\usepackage{epstopdf}
\usepackage{multicol}
\usepackage{stfloats}
\usepackage{url}
\usepackage{subcaption}
\usepackage{color}
\usepackage{enumerate}
\usepackage{comment}
\usepackage{breqn}
\usepackage{bbm}
\usepackage{cite}
\usepackage{caption}
\usepackage{enumitem}
\usepackage{array}
\ifCLASSINFOpdf
\else
\fi
\begin{document}

\title{mmWave Enabled Connected Autonomous Vehicles: A Use Case with V2V Cooperative Perception}

\author{Muhammad Baqer Mollah,~\IEEEmembership{Graduate Student Member,~IEEE,}
        Honggang Wang,~\IEEEmembership{Fellow,~IEEE,}
        \\ Mohammad Ataul Karim,~\IEEEmembership{Fellow,~IEEE,}
        Hua Fang,~\IEEEmembership{Senior Member,~IEEE}%

\thanks{(Corresponding Author: Honggang Wang)}
\thanks{Muhammad Baqer Mollah and Mohammad Ataul Karim are with the Department of Electrical and Computer Engineering, University of Massachusetts Dartmouth, MA 02747 USA (Emails: mmollah@umassd.edu; mkarim@umassd.edu).}

\thanks{Honggang Wang is with the Department of Graduate Computer Science and Engineering, Katz School of Science and Health, Yeshiva University, New York, USA (Email: honggang.wang@yu.edu).}
\thanks{Hua Fang is with the Department of Computer and Information Science, University of Massachusetts Dartmouth, MA 02747 USA (Email: hfang2@umassd.edu).}
\thanks{This work is supported by National Science Foundation (NSF) under Grant No. \# 2010366.}%
\thanks{This article has been accepted for publication in IEEE Network. This is the author's version which has not been fully edited and content may change prior to final publication.}
\thanks{DOI: https://dx.doi.org/10.1109/MNET.2023.3321520}
\thanks{\copyright 2023 IEEE. Personal use is permitted, but republication/redistribution requires IEEE permission. See https://www.ieee.org/publications/rights/index.html for more information.}
}

\markboth{IEEE Network, 2023}%
{Shell \MakeLowercase{\textit{et al.}}: Bare Demo of IEEEtran.cls for IEEE Journals}

\maketitle

\begin{abstract}
    Connected and autonomous vehicles (CAVs) will revolutionize tomorrow’s intelligent transportation systems, being considered promising to improve transportation safety, traffic efficiency, and mobility. In fact, envisioned use cases of CAVs demand very high throughput, lower latency, highly reliable communications, and precise positioning capabilities. The availability of a large spectrum at millimeter-wave (mmWave) band potentially promotes new specifications in spectrum technologies capable of supporting such service requirements. In this article, we specifically focus on how mmWave communications are being approached in vehicular standardization activities, CAVs use cases and deployment challenges in realizing the future fully connected settings. Finally, we also present a detailed performance assessment on mmWave-enabled vehicle-to-vehicle (V2V) cooperative perception as an example case study to show the impact of different configurations.
\end{abstract}


\IEEEpeerreviewmaketitle

\section{Introduction}
	\IEEEPARstart{T}{he} successful deployment of future connected transportation is dependent on the reliable and seamless operations of vehicular connectivity \cite{yang2021machine}. Employing advanced technologies will always pave the way toward next generation autonomous mobility by promoting connected and autonomous vehicles (CAVs). In particular, the connectivity of envisioned CAVs use cases will be entirely facilitated by communications vehicle-to-everything (V2X), enabling vehicle-to-vehicle (V2V), vehicle-to-infrastructure (V2I), vehicle-to-network (V2N), and vehicle-to-pedestrian (V2P), to name a few.
	
	Once fully connected, autonomous vehicles are more likely to have the ability to learn from surroundings (e.g., connected and non-connected vehicles, infrastructures) and make decisions accordingly after compiling the sensory data collected from own on-board units and neighbor vehicles. For instance, according to the report in \cite{2021Unlocking}, it is estimated that a single connected vehicle requires approximate total data rate for sensor data communications is 25 gigabytes per hour, and once fully connected and autonomous, it is expected to reach up to 500 gigabytes per hour. Likewise, automotive manufacturers are essentially focusing on extending their communications capability, while meeting the ever-growing demands from CAVs. On the other hand, millimeter-wave (mmWave), known also as the extremely high-frequency (EHF), band operates roughly from 30 GHz (10~mm) to 300 GHz (1~mm), allowing itself as a key enabler to facilitate this trend. Conversely, physical characteristics of mmWave band is different from that of traditional sub-6 GHz. Thus, to move forward with the progression of use cases of CAVs, it is essential to overcome their associated deployment challenges. Accordingly, we present in this article on how mmWave communications facilitate several opportunities and efficiently utilize the development of envisioned use cases of CAVs.

    Unlike \cite{li2022mobility, gu2022multimodality, bang2021millimeter}, where the authors mainly focus on mobility supports, MIMO beam selections, and algorithms for beam management on mmWave communications, respectively, we discuss both details and specifics on functionalities, use cases, and deployment challenges for mmWave enabled CAVs. More specifically, we first present the inherent propagation and channel characteristics of mmWave band before exploring other aspects. After that, we highlight most recent vehicular standardization activities by IEEE and 3GPP 5G NR, and we then focus on a number of prospective use cases enabled by mmWave assisted CAVs. Followed by the use cases, we investigate the potential challenges while CAVs are deployed at mmWave, including learning-based beamforming techniques, efficient resource management, authentication during frequent handovers, mmWave at 60 GHz, and real time deep learning in V2V. Before concluding, we validate the performances of mmWave enabled V2V cooperative perception and show the corresponding numerical insights.

\section{Background Features of mmWave}
    In this section, we outline the physical characteristics of mmWave and the necessity of beamforming techniques.
    
\subsection{Physical Characteristics of mmWave Band}
    The mmWave signals have unfavorable propagation characteristics specifically in outdoor environment. Compared to the signals at sub 6 GHz frequency bands, mmWave signals have a much higher propagation loss while travelling through the atmosphere. This is due to having small wavelength as comparable to the size of air molecules, mmWave signals face atmospheric attenuation with molecular absorption caused by oxygen ($\text{O}_{2}$), water vapor ($\text{H}_{2}\text{O}$), and other dust particles in the air. In general, the wireless communications based on mmWave bands are mostly dependent on the availability of line of sight (LoS) links and potentially viable for short distance communications.
    
    On the other hand, small wavelength of mmWave band essentially weaken the diffraction capability of the signals. In fact, because of this limitation, mmWave communications also suffer from a much higher sensitivity to blockage by vehicles, human, buildings, and tree like physical objects compared to lower carrier frequency band signals. Such blockage characteristics of mmWave propagation drastically limit the signal strength, which, may result in limiting the communications range. Nonetheless, there have been a number of considerable channel measurement efforts on mmWave frequency bands. For instance, the authors in \cite{rangan2014millimeter, sun2018propagation} conducted measurements at 28, 38, and 73 GHz with a focus on fully understanding the large-scale physical characteristics of these channels in urban outdoor environment. However, the results show that the mmWave bands are particularly suitable for communications within 100-200 m distance, even with non-line of sight (NLoS) links.

\subsection{Beamforming Techniques}
	Inherently, mmWave wireless links should be highly directional. The transmission loss and short distance communications at this high frequency band necessitate large-scale antenna arrays at both transmitter and receiver ends for realizing high directive gains. In fact, it is possible to configure many antenna arrays with small sizes due to their short wavelength. Using such large-scale antenna arrays is significant for beamforming, which aids the wireless links have an improved signal quality, i.e., signal-to-noise ratio (SNR), by accommodating stricter alignment between transmitter and receiver end beams guaranteeing a highly directional communication.
	
	Radio frequency-based beamforming, such as analog, digital, and hybrid techniques essentially have their own benefits and drawbacks depending on the application scenarios. Hybrid techniques, among them, aim to attain a balance between low-complexity analog and high-complexity fully digital techniques to obtain the benefits, including hardware costs and processing times from both analog and digital techniques. Perhaps, this hybrid beamforming, for example \cite{lota2022mmwave}, is particularly suitable for mmWave connected vehicular communications involving high mobility scenarios.
	
	Besides, the high directionality of mmWave communications imposes time overhead in longer beam tracking, selections, and alignment processes, thereby introducing higher latency. In fact, in order to address such latency issues, it is essential to employ adaptive beamforming techniques so that robust adaptation can be facilitated, which is one of the key steps to stabilize the wireless link quality while the vehicles are moving. As such, the authors in \cite{zhang2021adaptive} developed an adaptive beamforming technique for mmWave so that a large amount of sensory data can be disseminated among vehicles with multiple gigabit data rates and lower latency. In summary, because of the aforementioned unique characteristics, mmWave systems must rely heavily on large-scale antenna arrays, highly directional communications with beamforming techniques, and robust adaptive techniques to ensure reliable connections.

\section{Standardization Aspects}
    In this section, we highlight the standardization aspects in view of CAVs deployment and mmWave V2X for CAVs use cases.

\subsection{CAVs Deployment}
    European Telecommunications Standards Institute (ETSI), the International Organization for Standardization (ISO), the Society of Automative Engineers (SAE), and the American Association of State Highway and Transportation Officials (AASHTO) are actively developing standardization protocols related to CAVs deployment and ecosystem. Specifically, efforts for communicating standardized messages among the connected vehicles and surroundings are defined by ETSI \cite{charpentier2022proposal} into five types of messages, including (i) cooperative awareness notifications, (ii) decentralized environmental safety alerts, (iii) V2I, and vice versa, (iv) signal phase and timing information, and (v) map data. In this regard, the ISO is also involved in defining the communications for CAVs. For instance, the ISO/TS 19091\footnote{\url{https://www.iso.org/standard/73781.html}} present a standardized framework to establish the information exchange between roadside units (RSUs) and vehicles, specifying the data elements and structures to facilitate enhanced mobility and safety of the vehicles. 
    
    Further, the SAE International describes the functional approaches of message structure, encoding and decoding while implementing the 5.9 GHz Dedicated Short Range Communications (DSRC) standard in their J2735 Standard\footnote{\url{https://www.sae.org/standards/content/j2735set_202211/}}. However, this standard is designed particularly for DSRC messages; perhaps, it is possible to extend it by deploying along with other technologies as described in the J2735 standard as well. Nonetheless, SAE International also defines vehicle driving automation levels jointly with ISO\footnote{\url{https://www.sae.org/standards/content/j3016_202104/}}, which comprises six levels, such as no driving automation (level 0), assistance with human driver (level 1), automation partially (level 2), automation conditionally (level 3), automation highly (level 4), and automation fully (level 5). 
    
    Besides, the AASHTO specifies ten policy principles and various recommendations under these policy principles\footnote{\url{https://cav.transportation.org/}} for deployment as well as operations of CAVs. In particular, the policy principles cover (i) national strategy, (ii) safety, (iii) sustainability, (iv) future connectivity, (v) investment, (vi) quality of life, (vii) state and federal roles, (viii) uniform national policy, (ix) community engagement and collaboration with industry, and (x) security and privacy-preserving data sharing. 

\begin{figure*} [!b]
	\includegraphics[width=.7\linewidth]{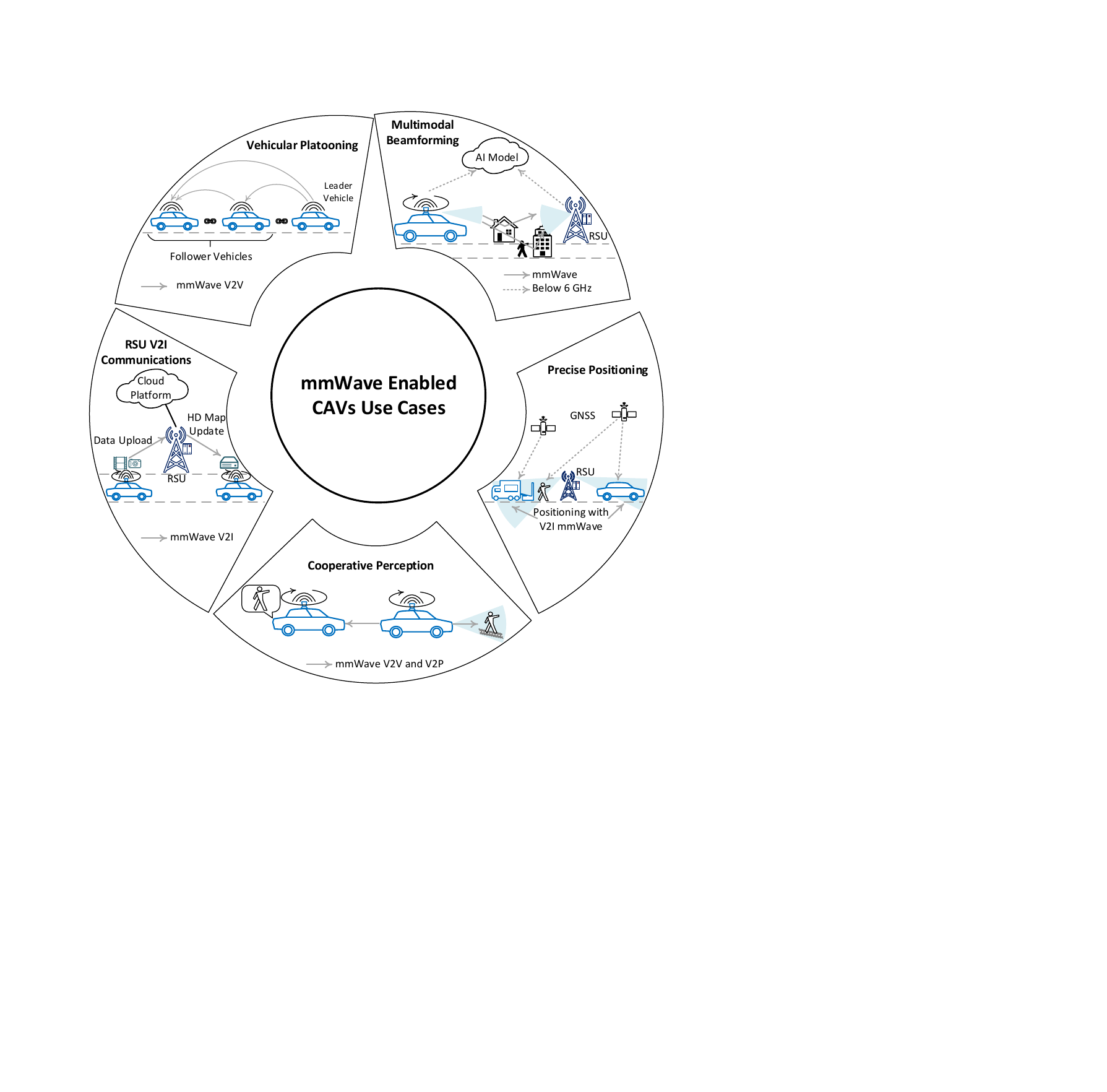}
    \centering
    \caption{Selective mmWave enabled use cases of CAVs.}
    \label{fig: Figure1}
\end{figure*}
        
\subsection{mmWave V2X Standardization Activities}
    3GPP 5G NR and IEEE 802.11 have already considered to integrate mmWave in V2X communications. While cellular V2X by 3GPP and IEEE 802.11p (DSRC) may have some limitations in terms of availability and reliability potentially demanded by CAVs use cases, both 3GPP and IEEE are currently promoting the standardization activities to address the limitations, such as 5G NR V2X \cite{misc1} and 802.11bd \cite{misc2}. These two standardization activities are presented as follows.

    \textit{3GPP 5G NR:} For the standardization activities, 3GPP 5G NR (New Radio) V2X and IEEE 802.11 NGV (Next Generation V2X) are the two key groups promoting the technical specifications and defining the advanced use cases. In general, 3GPP 5G NR has specified to operate in both sub-6 GHz as well as high band ($>$24 GHz, mmWave) frequencies. However, 3GPP standardization efforts in V2X arena have come up with C-V2X (cellular V2X) in Release 14 with long-term evolution (LTE) in two parts. The first part, referred as Direct Communications, has been aimed to provide typical safety related services within V2X. Whereas, the second part serves as cellular network to provide services which require wide area coverage, thus, named as Vehicle-to-Network. Later, Releases 15 and 16 described the native supports for mmWave band by enabling improved localization accuracy of the vehicles, employing large-scale antenna arrays, and exploiting extra spatial and angular degrees of freedom within 5G New Radio (NR) carrier. Further, 3GPP’s most recent release 17 \cite{misc1} has focused on supporting enhanced V2X communications within mmWave band by including resource allocations techniques in multi-casting, user equipment (UE) power savings, and improved sidelink for V2V operations. This release 17 has also introduced a channel model at frequency range 2 (FR2), which, in principle, covers from 24.25 GHz to 52.6 GHz. Clearly, the FR2 has a large amount of spectrum, and consequently, it's technical specifications will help to move forward the mmWave V2X communications to enable ultra-low-latency services. In fact, it is expected that the utilization of other bands will be considered to include soon under other releases since the demands for more 5G NR applications are becoming increasingly significant in autonomous driving field.
    
    \textit{IEEE 802.11 NGV:} Similarly does, IEEE 802.11 NGV was initiated to develop IEEE 802.11 specifications in physical (PHY) and medium access control (MAC) layers particularly for V2X communications. Thereby, the first amendment building upon IEEE 802.11 was IEEE 802.22p to offer wireless access in vehicular environment (WAVE), which presented new functionalities to provide inter-vehicle connectivity to the vehicles in dynamic scenarios within DSRC. While enabling inter-vehicles communication was a significant achievement, IEEE 802.11p WAVE is unable to fulfill the low latency and high throughput requirements of recent use cases in semi or fully autonomous driving, such as cooperative driving, High-Definition (HD) map updating in real-time, and vehicular platooning. For this reason, inherited from IEEE 802.11p, the IEEE 802.11 NGV has recently developed IEEE 802.11bd PHY and MAC specifications in \cite{misc2} with a focus on 5.9 GHz band, but also support from 57 GHz to 71 GHz, which is considered the first standard using new spectrum at mmWave band for V2X communications. Basically, this IEEE 802.11bd standard has redefined fundamental principles of PHY and MAC layers through incorporating innovative techniques to take benefits of beamforming gains and cope with mmWave propagation losses. Besides, this standard has also defined to support increased throughput, reduced end-to-end latency, and higher channel bandwidth up to 40 MHz, while ensuring coexistence and backward compatibility with IEEE 802.11p, among other advancements. However, apart from this IEEE 802.11bd, IEEE 802.11 has two more specifications on mmWave, namely IEEE 802.11ad and IEEE 802.11ay, but these two standards are targeted to serve indoor wireless communications applications.
    
    To sum up, both 3GPP 5G NR-V2X and IEEE 802.11bd represent the latest mmWave enabled specifications for V2X employing innovative techniques and procedures, which could be considered as basis of opening up new possibilities for V2X communications in the upcoming beyond 5G and 6G era, thereby, have a vision on fulfilling the requirements of advanced use cases of future fully connected vehicles.

\section{mmWave in CAVs Use Cases}
    Once CAVs are fully deployed, mmWave communications is expected to support a number of use cases. In the following, we discuss such CAVs use cases, including but not limited to.
    
    \subsection{Cooperative Perception}
    It is possible to share the perceptual data in cooperative driving, for example, LiDAR sensor data between vehicles and process them for the purpose of avoiding potential collisions. In the context of CAVs, referred to as cooperative perception, and it is particularly suitable for use in complex urban areas. In fact, cooperative perception techniques facilitate autonomous vehicles to enlarge their visibility areas by developing a layout of surrounding objects and traffic, such as pedestrians, bicycles, buildings, and non-connected vehicles. With this functionality, indeed, the vehicles can collect the perceptual data from connected neighboring vehicles through V2V communications in order to manage blind spots, detect hidden objects, and recognize NLoS traffic ahead. However, for this use case, leveraging mmWave is necessary, specifically for real-time integration of multimedia data sharing and processing. Both high data rate, lower latency, and highly reliable communications are desired, thereby ensuring safe automated driving.
    
    \subsection{Vehicular Platooning}
    Instead of driving individually, a group of vehicles may consider driving together by making a platoon on highways, which results in potentially lower fuel consumption, reduced congestion, increased lane capacity as well as improved safety, eventually leading to optimized transportation. In this platoon formation, the lead vehicle may have a driver, whereas the other following vehicles can be driverless, and the lead vehicle basically represents the speed and trajectory reference for the followers. However, in such a connected driving concept, V2V communications are expected to play an important role in maintaining connectivity and exchanging the high-volume of onboard sensor data among the vehicles. Besides, the vehicles in platoon travel maintain relatively shorter spacing in-between. Essentially, compared to DSRC, V2V enabled by mmWave has become one of the most suitable communication technologies to enable vehicle platooning.
    
    \subsection{Precise Positioning}
    The unique characteristics of mmWave band, such as having large arrays of small sized antennas for directional communications, can be employed to provide highly accurate positioning for automated driving, which is considered as one of the key requirements for future CAVs. Therefore, mmWave enabled CAVs is expected to possess the capability of precise positioning, which is able to offer environmental perception in real-time or near-instantly. However, mmWave along with global navigation satellite system (GNSS) and time of arrival (ToA) type localization techniques are utilized to improve the positioning accuracy, perhaps most importantly, by employing multiple techniques to allow for centimeter level accuracy positioning. Such capabilities lead to more efficient and safe automated driving, particularly in urban vehicular scenarios. In fact, native support for mmWave applicable positioning has been introduced in Release 16 of 3GPP 5G NR.
	
    \subsection{Multimodal Beamforming}
    To increase the performance of beamforming in terms of reduced processing time and improved directional accuracy, an enhanced beamforming method that uses GPS, Camera, Radar, and LiDAR like multiple modalities from both non-RF data sources can be utilized along with RF-only mmWave can be employed. This beamforming method is referred as multimodal beamforming. It can be used to detect pedestrians and any other objects either between two connected vehicles or between a vehicle and a roadside unit. Accordingly, deep learning-assisted object detection techniques are applied after fusing the non-RF visual sensor data to develop a surrounding visualization of the vehicle. For instance, the work in \cite{salehi2022deep} presents a beam selection procedure utilizing deep learning on distributed multimodal data. The technique when employed to process synthetic and home-grown real-world data sets has shown to yield 95\% and 96\% enhancement respectively beam selection speed as well as predicting the top-10 best beam pairs.
    
	\subsection{RSU V2I Communications}
	Having a dynamic high-definition 3D map is essential in automated driving to maneuver autonomous vehicles efficiently and safely. The onboard visual sensors are used to build crowdsourced maps presented as real-time point clouds and high-resolution multimedia data. However, due to the necessity of keeping the map updated in real-time, the vehicles are essentially dependent on the RSUs by considering edge computing nodes, which may introduce a significant delay while collecting and processing the necessary information. Ensuring the required high throughput as well as stringent latency with mmWave V2I can help to enhance the capability of serving on-demand requests on location-aware and time-sensitive updates, navigation and control by obtaining the perception of the surroundings instantly, and providing, in turn, safe operation in automated driving.
	
    In summary, these aforementioned CAVs use cases, shown in Fig. \ref{fig: Figure1}, necessitate adherence to the strict requirements given very high data rate, lower latency, precise positioning, and highly reliable communications. And to satisfy these demands, CAVs' use cases promise to be highly beneficial in mmWave communications. Accordingly, connectivity with mmWave communications is becoming an increasingly important topic in the CAVs areas.
    
\begin{figure*} [ht]
	\includegraphics[width=.9\linewidth]{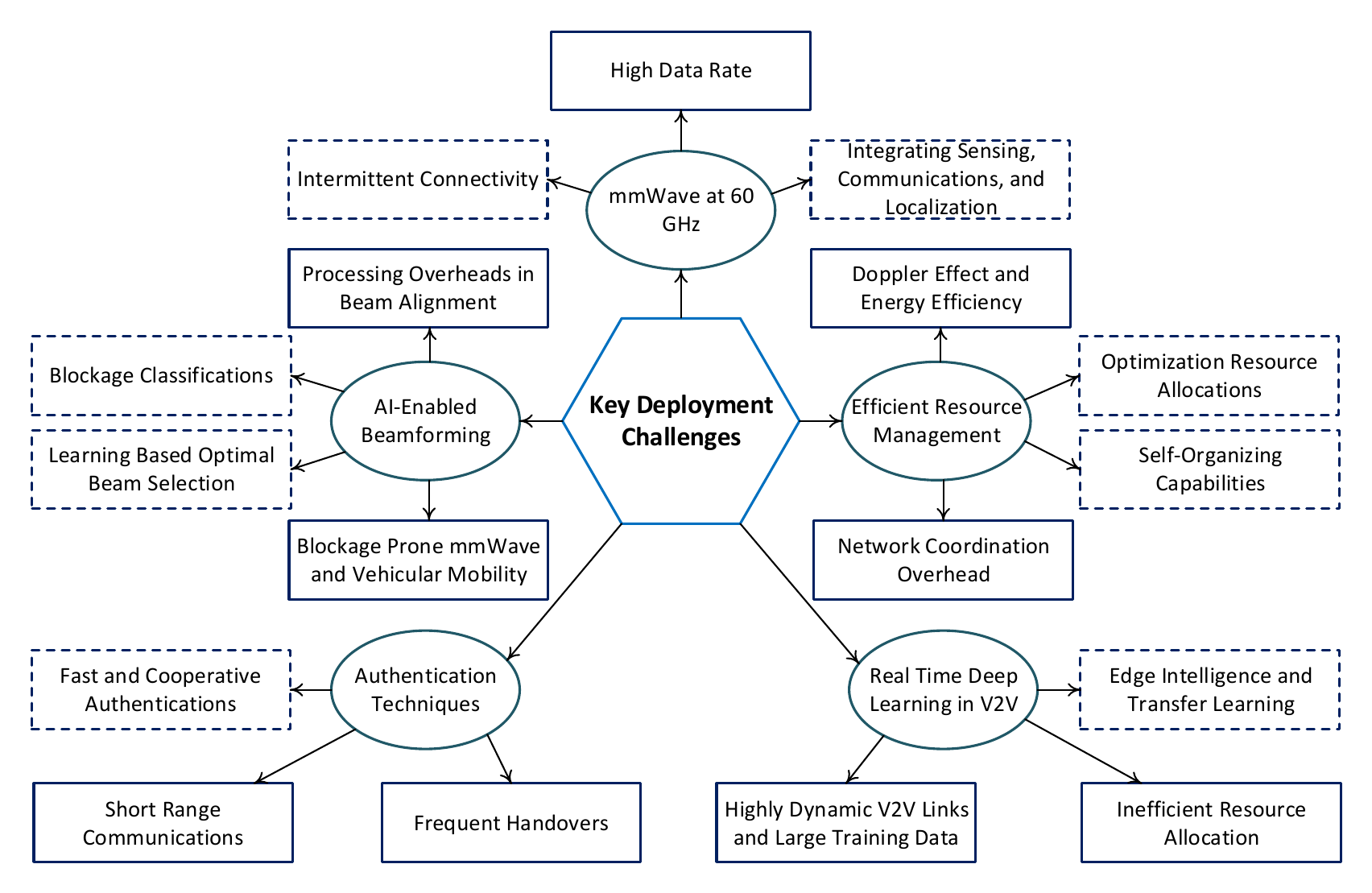}
	\centering
	\caption{A summary of key deployment challenges of mmWave enabled CAVs.}
    \label{fig: Figure2}
\end{figure*}

\section{Deployment Challenges}
In this section, we discuss key challenges that deserve further attention while deploying mmWave enabled advanced CAVs use cases as follows with Fig. \ref{fig: Figure2} representing a summary of such challenges.
    
    \textit{AI-Enabled Beamforming:} Blockage-prone mmWave communications introduce significant challenges in beam alignment algorithms while maintaining reliable transmitter and receiver links in high mobility and dynamic environments, which can severely impact delay-sensitive CAVs. Consequently, such beam alignment algorithms are required to perform adaptive and frequent updates of large beamforming arrays, perhaps adding further processing overheads. However, facilitated with the capability of building the trained model from calibrated beam information and RF signals, it is possible to utilize the trained model later to predict the optimal beam selection, eventually, realizing higher beamforming gain. Likewise, learning-based techniques can also be utilized for the purpose of blockage classification and afterward, predict the beam pairs to bypass the blockage, resulting in maximizing the mmWave network capacity. For example, the authors in \cite {al2022intelligent} present to equip the mmWave base stations with RGB cameras to get the surrounding sensing information, such as moving vehicles, pedestrians, etc. Through utilizing the multivariate regression and vision-based algorithms on the surrounding sensing information, the base stations can proactively predict the potential LoS link blockages, thereby avoiding sudden link failures by doing timely handover. 
    
    \textit{Efficient Resource Management:} Doppler effect caused by vehicular mobility impacts the energy efficiency and network capacity of mmWave enabled CAVs. Accordingly, the network coordination overhead costs increase essentially with the increase of desired high data rates. These in turn, however, demand to reconsider developing effective resource management as well as optimization techniques, such as network planning, optimized allocation of resources (e.g., bandwidth, channel, power), admission control, and scheduling. Moreover, promising self-organizing networks could support developing dynamic network configuration and planning. In particular, such self-organizing capabilities should also consider the sensitivity of specific CAVs use cases to include complex interference, routing, and time delay caused by vehicular mobility and mmWave communications.
    
    \textit{Authentication during Frequent Handovers:} Due to short range communication capability, connected vehicles enabled by mmWave communications will require frequent handover while moving. In mmWave network, a part of security mainly relies on authentication techniques for both messages and involved parties. However, frequent handovers essentially demand multiple authentications, which may result in unexpected communications overheads. To be effective, fast and cooperative authentication techniques need to be further considered in the future while realizing successful authentications with moderate communication costs.

\begin{figure*} [ht!]
\begin{subfigure}[b]{0.30\textwidth}
	\centering
	\includegraphics[width=5.8cm]{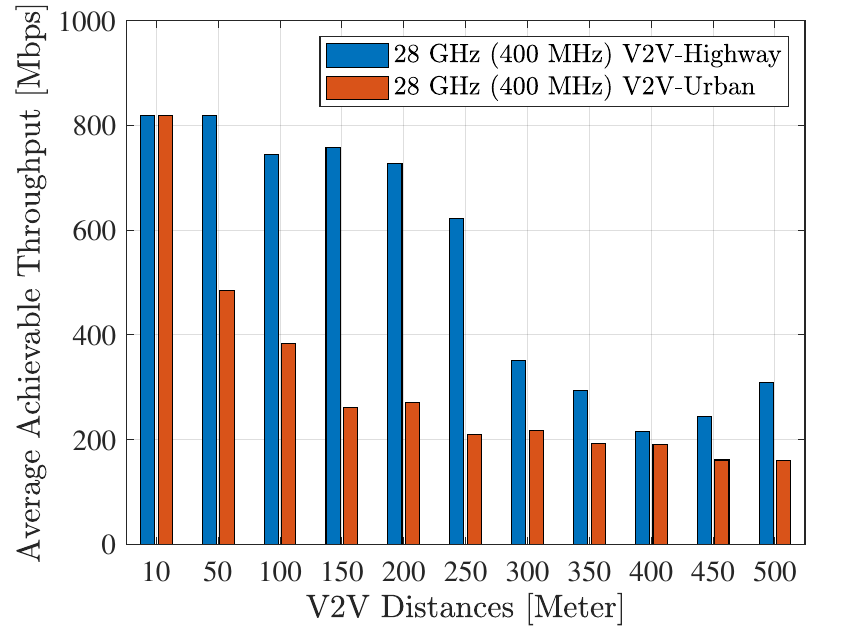}
    \caption{\centering \scriptsize Throughput performances in 28 GHz with highway and urban scenarios}
\end{subfigure}
    \hfill
\begin{subfigure}[b]{0.30\textwidth}
	\centering
    \includegraphics[width=5.8cm]{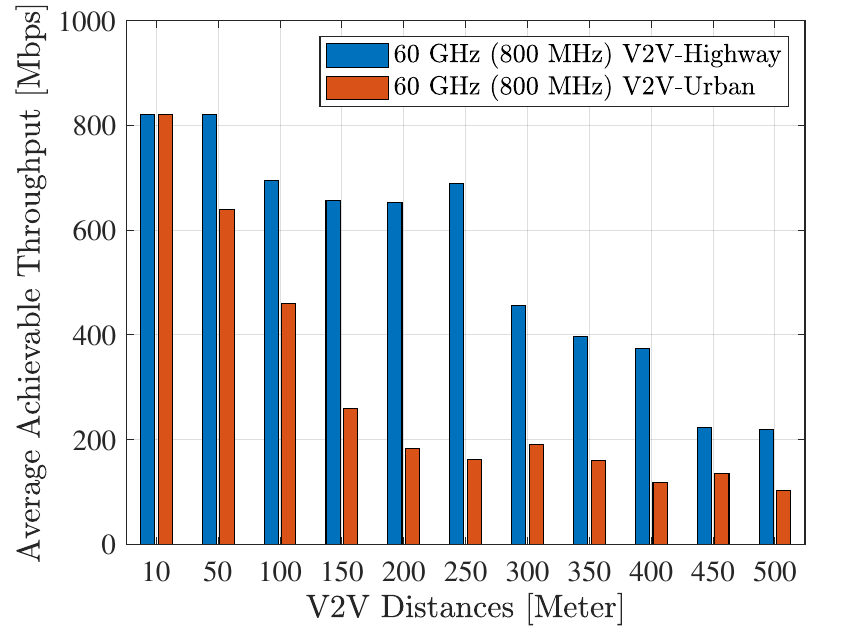}
    \caption{\centering \scriptsize Throughput performances in 60 GHz with highway and urban scenarios}
\end{subfigure}
	\hfill
\begin{subfigure}[b]{0.30\textwidth}
    \centering
	\includegraphics[width=5.8cm]{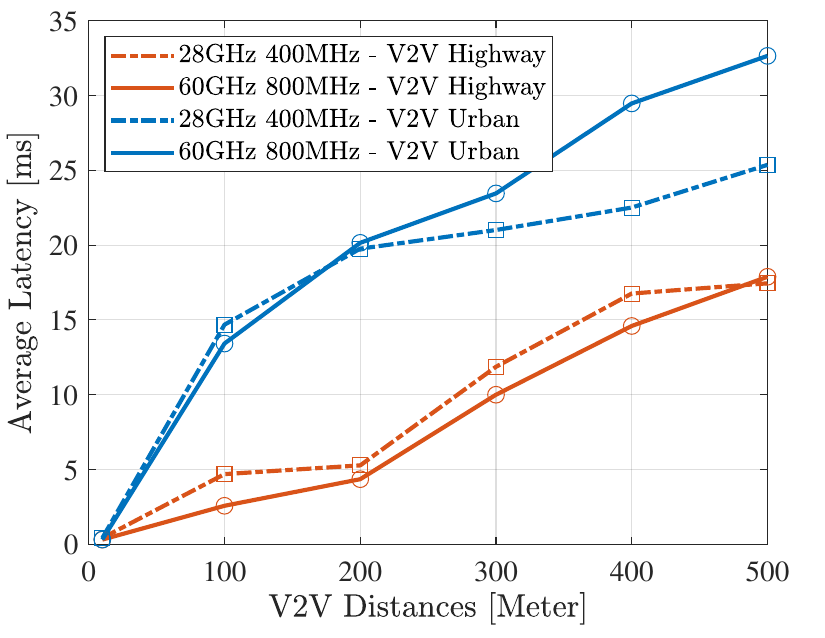}
    \caption{\centering \scriptsize Latency performances in 28 GHz and 60 GHz with highway and urban scenarios}
\end{subfigure}
	\caption{(a)\&(b) The impacts of V2V distances on achievable throughput and (c) Results of latency performances with respect of V2V distances in mmWave communications at 28 GHz and 60 GHz with two propagation scenarios.}
	\label{fig: Troughput_Latency}
\end{figure*}
    
    \textit{mmWave at 60 GHz:} In recent years, mostly lower bands of mmWave have been studied well, for example, at 28 GHz and proven the feasibility of employing V2X communications. However, mmWave at unlicensed 60 GHz (57 – 64 GHz) has also been considered as a promising frequency band for the upcoming paradigm shift on the procedures of communications technologies utilized to support specifically the growing high data rates demands from V2X for CAVs. On the other hand, to make highly directional communications through employing stricter alignment between receiver and transmitter end beams inherently brings its own unique limitations. Thus, it is necessary to do feasibility studies of CAVs in 60 GHz communications, including investigating the intermittent connectivity among vehicles, RSUs, and other entities. We believe that reducing the impact of the Doppler effect is necessary for high data rate (e.g., gigabyte) applications. Besides, incorporating sensing, communications, and localization like different functionalities along with other mmWave and below 6 GHz spectrum can be a promising direction to facilitate development of realistic use cases.
  
    \textit{Real Time Deep Learning in V2V:} It is worth mentioning that deep learning techniques have attracted growing interest in mmWave enabled vehicular communications because of having capability of enhancing the performances of communications. Through training the ray-tracing (e.g., angle of arrival, angle of departure, time of arrival, received power), channel state information, LiDAR, RADAR, camera, and location data, deep learning techniques have made possible to extract the features of the system, which eventually result improved vision-based applications for connected vehicles. On the other hand, when it comes to CAVs uses cases having V2V communications, such as collaborative perception and platooning, inefficient resource allocation (e.g., channel, energy budget, bandwidth) techniques and dynamic V2V links may lead performance degradation substantially while deploying real-time deep learning algorithms. Besides, it becomes more challenging when the vehicles have to handle a large amounts of data for training. However, edge intelligence algorithms can be utilized to alleviate the high communications overhead, latency, and complexity, and also, transfer learning techniques can be applied so that the trained models or knowledge can be effectively reused. 
    
\begin{figure*} [!b]
\begin{subfigure}[b]{0.50\textwidth}
	\centering
	\includegraphics[width=7cm]{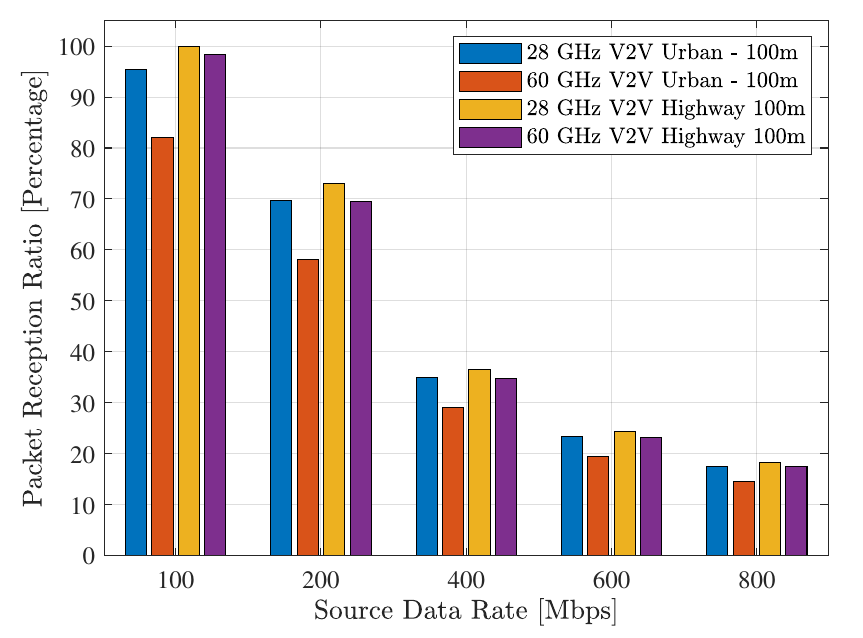}
	\caption{\centering \scriptsize 28 GHz and 60 GHz with highway and urban scenarios in 100-meter distance}
\end{subfigure}
\begin{subfigure}[b]{0.50\textwidth}
    \centering
    \includegraphics[width=7cm]{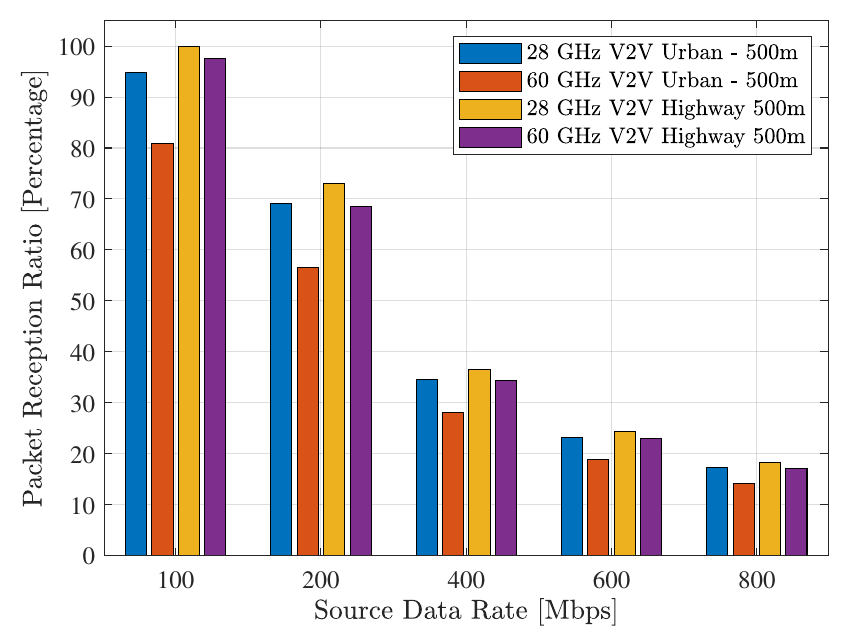}
	\caption{\centering \scriptsize 28 GHz and 60 GHz with highway and urban scenarios in 500-meter distance}
\end{subfigure}
	\caption{The impact of packet reception ratio over source data rate. The packet reception ratio is significantly becoming smaller with source data rate decreasing.}
	\label{fig: PRR}
\end{figure*}

\section{A Use Case: V2V mmWave Communications}
    In this section, we present a detailed performance study of a practical CAVs use case to realize how high frequency mmWave bands impact various aspects of vehicular scenarios. However, due to the costly mmWave hardware and the complexity of developing as well as testing the network components, it might be impractical to validate the performance matrices with a fully implemented testbed. For this reason, we choose MilliCar presented in \cite{drago2020millicar} to perform the end-to-end simulations, which is a Network Simulator-3 (NS-3) mmWave module followed by the NS-3 mmWave module by NYU-Wireless\footnote{\url{https://github.com/nyuwireless-unipd/ns3-mmwave}}. On the other hand, we choose V2V cooperative perception communications at mmWave bands as an example use case to implement and analyze the corresponding performances. In particular, cooperative perception requires perceptual data sharing with very high throughput within short distance, which makes mmWave V2V communications as the most suitable use case. While the best mmWave in CAVs use cases may not be just limited to V2V cooperative perception, the parameters in the simulation environment are chosen in accordance with the V2V cooperative perception use case. 
    
    The simulated settings are comprised of two representative scenarios, namely vehicular highway and urban scenarios having medium traffic density, with highway covering 1500 vehicles/hour/direction and urban area covering 316 vehicles/$\text{mile}^2$. Both scenarios may include LoS, Non-LoS (NLoS), and Vehicle NLoS (NLoSv) types of states depending on the vehicle moving directions and surroundings. We perform the simulations on V2V links operating at $f_c$ = 28 GHz (with 400 MHz bandwidth) and 60 GHz (with 800 MHz bandwidth) frequencies, with numerology index 3 in the physical layer (sub-carrier spacing = 120 kHz and slots per frame = 8) and the vehicles exchange 1024 bytes of User Datagram Protocol (UDP) packets with each other by NR sidelink interface in order to share the perceptual data unless otherwise mentioned. Accordingly, the obtained simulation results are illustrated in Fig. \ref{fig: Troughput_Latency} and Fig. \ref{fig: PRR}.
    
    Figs. \ref{fig: Troughput_Latency}a and \ref{fig: Troughput_Latency}b highlight the achievable average throughput and latency, respectively, over different V2V distances ($\mathcal{D}$ = 10 meters to 500 meters) while two vehicles are moving one behind the other in the same lane at approximate 45 mph speed. From the results, we can observe the following points.
    
\begin{itemize}
    \item As can be noticed, mmWave at 28 GHz and 60 GHz highway scenarios have higher throughput and lower end-to-end latency than the urban scenarios up to a certain V2V distance ($\sim$300 m). This is due to the obstacle free signal propagation in highways, resulting in the high probability of LoS.
    
    \item Because of significant path loss, the achievable throughput of mmWave links significantly drops as the V2V distance increases. At the 350 meters, even the LoS V2V links is very likely to be blocked in highway scenarios. Thereby the throughput starts becoming worse.
    
    \item Presence of many obstacles leads to high penetration loss in urban scenarios can be drastic and subsequently provide the low performance of achievable throughput. We can observe it specifically in urban cases by comparing the bar lines from V2V distances at 300m and 200m corresponding to the 28 GHz and 60 GHz, respectively. However, proper beamforming management techniques may compensate partially or fully for it.
    
    \item The size of the buffer in this simulation setting is fixed at 500 packets. Although getting high throughput depends on the buffering, large buffer size may result in increased end-to-end delay because of buffering at the queue.
    
\end{itemize}

    Figs. \ref{fig: PRR}a and \ref{fig: PRR}b depict the impact of Packet Reception Ratio (PRR) over different source data rate values at 28 GHz and 60 GHz. In this simulation setting, we consider there are four vehicles divided into two groups moving on the same lane at approximately 45 mph speed. We also consider each link end is equipped with 4 antenna elements, the resources are shared orthogonally among the vehicles, and 64-QAM modulation technique is employed. With these settings, we can get a notable increase in PRR values over a lower source data rate by comparing the curves. Specifically, by comparing the curves, the high frequency at 60 GHz has slightly lower PRR values than 28 GHz. This is again due to mmWave signal attenuation problems. Indeed, with a lower source data rate, the PRR value is close to 1, which is very suitable to safety-critical services. However, robust forward error correction (FEC) techniques might be required to maintain high PRR.
    
\section{Conclusion}
   In this article, we have presented a detailed discussion on the prospects of mmWave on envisioned CAVs use cases. Specifically, we have detailed the physical characteristics of mmWave signals and the necessity of beamforming techniques. We also have discussed the mmWave standardization activities, which apply to CAVs specified by IEEE and 3GPP 5G NR. After that, we have highlighted the selected use cases of CAVs and discussed how and to what extent mmWave can support them. Further, we have investigated several specific deployment challenges. Finally, we have evaluated the performance of a V2V cooperative perception scenario as an example use case. However, we plan to extend the performance evaluation part to prototype development and then evaluate the performance in our future works. We expect this article will facilitate further exploration in this crucial research area.

\section*{Acknowledgments}
    The authors would like to thank the editor and anonymous reviewers for their constructive suggestions and valuable comments.

\ifCLASSOPTIONcaptionsoff
  \newpage
\fi

\bibliographystyle{IEEEtran}
\bibliography{bibliography.bib}

\hfill \break
    
    \begin{center}
    \textbf{Biographies}
    \end{center}
    
    \textbf{Muhammad Baqer Mollah} is currently pursuing a PhD degree in the Department of Electrical and Computer Engineering at the University of Massachusetts Dartmouth. His research interests include advanced communications, security, and edge intelligence techniques for Internet of Things (IoT) and connected vehicles.
    \hfill \break
    
    \textbf{Honggang Wang} is a Professor in Department of Graduate Computer Science and Engineering at Katz School of Science and Health, Yeshiva University. His research interests include AI and Internet of Things, connected vehicles, smart health, and cyber security. He served as the Editor-in-Chief of IEEE Internet of Things Journal (2020 - 2022). 
    \hfill \break

    \textbf{Mohammad Ataul Karim} is a Professor in the Department of Electrical and Computer Engineering at the University of Massachusetts Dartmouth. His areas of research encompass optical computing, information processing, pattern/target recognition, computer vision, and Internet of Things.
    \hfill \break
    
    \textbf{Hua Fang} is a Professor in the Department of Computer and Information Science at the University of Massachusetts Dartmouth. Her current research interests involve real-time machine or statistical learning, and visual analytics of wearable biosensor data streams, broadly in E-/M-/Digital/Tele-/Virtual/Precision health, and Internet of Things.

\end{document}